\documentclass[11pt]{article}
\usepackage{amssymb}
\usepackage{amsthm}
\usepackage{amsmath}
\usepackage[dvips]{epsfig}
\usepackage{graphicx}
\usepackage{color}
\usepackage{url}

\begin{document}

\title{\bf Nonlinear dynamics in PT-symmetric lattices}

\author{Panayotis G. Kevrekidis$^{1}$, Dmitry E. Pelinovsky$^{2,3}$, and Dmitry Y.Tyugin$^{3}$ \\
$^{1}$ {\small \it Department of of Mathematics and Statistics, University of
Massachusetts, Amherst, MA 01003-9305, USA} \\
$^{2}$ {\small \it Department of Mathematics and Statistics, McMaster
University, Hamilton, Ontario, Canada, L8S 4K1 } \\
$^{3}$ {\small \it Department of Applied Mathematics, Nizhny Novgorod State
Technical University, Nizhny Novgorod, Russia }}

\date{\today}
\maketitle

\begin{abstract}
We consider nonlinear dynamics in a finite parity-time-symmetric chain
of the discrete nonlinear Schr{\"o}dinger (dNLS) type. We work in the
range of the gain and loss coefficient when the zero equilibrium state
is neutrally stable. We prove that the solutions of the dNLS equation
do not blow up in a finite time and the trajectories starting with
small initial data remain bounded for all times.
Nevertheless, for arbitrary values of the gain and loss parameter,
there exist trajectories starting with large initial data that grow
exponentially fast for larger times with a rate that is rigorously
identified. Numerical computations illustrate
these analytical results for dimers and quadrimers.
\end{abstract}

\section{Introduction}

Considerable recent interest in the physical literature have been
devoted
 to lattices of
the discrete nonlinear Schr\"{o}dinger (dNLS) type with compensated gains
and losses.
For the one-dimensional lattice, a prototypical, mono-parametric
model takes the form
\begin{equation}
\label{dnls}
i \frac{d u_n}{d t} = u_{n+1} - 2 u_n + u_{n-1} + i \gamma (-1)^n u_n + |u_n|^2 u_n,
\end{equation}
where parameter $\gamma$ stands for the gain and loss coefficient. If the lattice is truncated at
a finite chain, then the site index $n$ ranges from $1$ to $2N$ for a positive integer $N$
subject to the homogeneous Dirichlet boundary conditions $u_0 = u_{2N+1} = 0$.

The dNLS model (\ref{dnls}) represents one of the simplest discrete systems
which are symmetric with respect to combined parity (P) and time-reversal (T)
transformations. Hence, it is referred to as the PT-dNLS equation.
Motivated by progress in optics both at the theoretical~\cite{Kuleshov,ziad,Muga} and
experimental~\cite{salamo,dncnat} levels, many researchers have studied
nonlinear stationary states of few site configurations \cite{Dmitriev,Li,Guenter,Ramezani,suchkov,Sukhorukov,ZK}
as well as in infinite lattices \cite{Pelin1,Sukh,zheng}.

In what follows, we will be particularly interested in dimers and 
quadrimers with alternating loss and gain in the adjacent sites.
It should be noted that such configurations are directly amenable
to the experimental considerations in \cite{dncnat}.
In fact, in such finite waveguide arrays, due to the presence of
iron doping, there
unavoidably exists some loss in all the channels (lattice sites).
However, the experimental procedure uses a mask covering
some of the waveguides~\cite{dkip}. Then the ones
that are {\it not} covered from the top are optically
pumped and experience gain. There is considerable freedom
in the selection of the mask so different configurations
(such as loss, loss, gain, gain, or loss, gain, loss, gain)
are feasible. However, the key feature is that because
of the geometric characteristics of the waveguide
(with each channel being about 4 $\mu$m wide,
but 30 mm long), only a homogeneous pump beam can be
practically applied to all the gain channels, as in the
case considered hereafter. Notice, however, that our analysis
will be generalized beyond these configurations
to arbitrary finite chains.

Our previous work \cite{Pelin3} was devoted to the systematic analysis of nonlinear stationary states
in the finite PT-symmetric chains. In particular,
working in the range of the gain and loss coefficient $\gamma$
when the zero equilibrium state is neutrally stable \cite{Bar,Pelin2},
we continued stationary states from the two limits characterized by small and large
amplitudes of the stationary states. For these continuations, we modified the
arguments of the Lyapunov--Schmidt reduction method to establish the
existence of the solutions.
We have also illustrated numerically that the stationary states are stable
for small amplitudes and some of them are stable even for large amplitudes.

The nonlinear dynamics of oscillators with alternating loss and 
gain is an interesting problem in its own right.
Aspects of such dynamics that have been previously explored include
the following.
Transmission resonances in a PT-symmetric dimer coupled with a linear dNLS chain were studied
in \cite{Mirosh}. Asymmetric wave propagation through a finite PT-symmetric chain
was observed in \cite{lepri}. Nonlinear dynamics of wave packets near the phase transition
point was addressed with the reduction to the nonlinear Klein--Gordon equation in \cite{nixon1,nixon2}.

Our present paper studies the nonlinear dynamics of oscillators in the simplest finite
PT-symmetric chains such as dimers and quadrimers. We shall prove the following three main results
by using methods of the theory of differential equations.
Although (most of) these results are expected on an intuitive level, the rigorous proof of these results
appears to be a mathematical question of interest in its own right
and can be seen, in some of the cases below, to present
significant challenges. The three main
results are summarized as follows:

\begin{itemize}
\item[R1] We prove that solutions of the PT-dNLS equation do not blow up in a finite time.
This is achieved by a priori energy estimates for squared amplitudes of the nonlinear oscillators.
In the Hamiltonian case $\gamma = 0$, this result is equivalent to the conservation of the total
sum of squared amplitudes, so the calculation proceeds in a fashion parallel
to the corresponding conservation law and bears an extra step associated with Gronwall's lemma.

\item[R2] For parameter values of the gain and loss coefficient $\gamma$
when the zero equilibrium state is neutrally stable (this regime is referred to as the {\it exact PT-phase}), we prove
that the solutions of the finite PT-dNLS equation starting with small initial data remain bounded for all times.
This is expected in Hamiltonian systems with the energy conservation when the zero equilibrium
is stable. Nevertheless, the PT-dNLS equation with $\gamma \neq 0$
is non-Hamiltonian and lacks generally any conserved quantities.

\item[R3] For the same values of $\gamma$, we prove that despite the fact
that the zero equilibrium state is neutrally stable, there
exist solutions of the finite PT-dNLS equation starting with sufficiently large initial data which grow exponentially
fast for larger times. This result is perhaps the least expected among the three, because the values of $\gamma$
are inside the exact PT-phase, yet it can be anticipated since the balance between damped and gained
oscillators is broken at large initial amplitudes, hence the amplitude of the oscillators with gain grows
exponentially according to the linear law, in spite of the contributions of the nonlinear terms and
interactions with other damped oscillators.
\end{itemize}

The article is structured as follows. Section 2 gives the rigorous proof of the three main results
for the dimer. Although the dimer equations are fully integrable because of
the presence of conserved quantities (reviewed in Appendix A), we develop
qualitative methods of analysis, which become suitable for more complicated
finite PT-symmetric chains. Section 3 outlines the proof of the three main results
for the general case of finite PT-symmetric chains. Section 4 illustrates the main results
 with numerical computations of dimers and quadrimers. Section 5 contains
a summarizing discussion and some consideration of future challenges.

\vspace{0.5cm}

{\bf Acknowledgments:} The work of P.K. is partially supported by the US
National Science Foundation under grants NSF-DMS-0806762, NSF-CMMI-1000337,
and by the US AFOSR under grant FA9550-12-1-0332.
The work of D.P. and D.T. is supported by the ministry of education
and science of Russian Federation (Project 14.B37.21.0868).

\section{Nonlinear dynamics of a dimer}

Setting $a := u_1$ and $b := u_2$ for $N = 1$,
we consider the following system of two differential equations describing
a PT-symmetric dimer:
\begin{eqnarray}
\left\{ \begin{array}{l} i \frac{da}{dt} = b - i\gamma a + |a|^2 a, \\
i \frac{d b}{dt} = a + i \gamma b + |b|^2 b. \end{array} \right.
\label{dimer}
\end{eqnarray}
We fix the gain and loss parameter $\gamma$ in the interval $(0,1)$, which guarantees
neutral stability of the zero equilibrium; this is the parametric interval
of the exact PT-phase for this model. Indeed, for any $\gamma \in (0,1)$,
the zero equilibrium point is neutrally stable with the eigenvalue-eigenvector pairs:
\begin{eqnarray}
\label{eigenvalues-dimes}
\lambda = \pm i \sqrt{1 - \gamma^2}, \quad \left( \begin{array}{c} a \\ b \end{array} \right) =
\left( \begin{array}{c} 1  \\ i \gamma \mp \sqrt{1 - \gamma^2} \end{array} \right).
\end{eqnarray}
The system of dimer equations (\ref{dimer}) is fully integrable with two conserved quantities for
any $\gamma \neq 0$. As a result, the three main results R1--R3 can be proven with the use of conserved
quantities, as is outlined in Appendix A. Because the conserved quantities do not seem to exist
for more complicated PT-symmetric configurations, such as quadrimers, we shall prove the
same results using more general
qualitative methods of the differential equation theory.

\vspace{0.25cm}

{\bf Proof of R1:} It follows from the system of dimer equations (\ref{dimer}) that
\begin{eqnarray}
\left\{ \begin{array}{l} \frac{d |a|^2}{d t} = - 2\gamma |a|^2 + i( \bar{b} a - b \bar{a}), \\
 \frac{d |b|^2}{d t} = 2\gamma |b|^2 - i( \bar{b} a - b \bar{a}). \end{array} \right.
\label{dimer-amplitudes}
\end{eqnarray}
Adding these two equations together, we obtain the balance equations for squared amplitudes
\begin{equation}
\label{balance-dimer}
\frac{d}{d t} ( |a|^2 +  |b|^2) = 2\gamma ( |b|^2 - |a|^2).
\end{equation}
By Gronwall' inequality, the balance equation (\ref{balance-dimer}) results
in the a priori bound
\begin{equation}
\label{Gronwall}
|a(t)|^2 + |b(t)|^2 \leq (|a_0|^2 + |b_0|^2) e^{2 \gamma t}, \quad t \in \mathbb{R},
\end{equation}
where $a_0 = a(0)$ and $b_0 = b(0)$. A priori bound (\ref{Gronwall}) proves that
the amplitudes $|a(t)|$ and $|b(t)|$ do not blow up in a finite time.

\vspace{0.25cm}

{\bf Proof of R2:} The key point of the proof is to rewrite the system of differential equations
(\ref{dimer-amplitudes}) in the integral form:
\begin{equation}
\label{integral-amplitude-a}
|a(t)|^2 = |a_0|^2 e^{-2 \gamma t} + i \int_{0}^t e^{-2 \gamma (t-\tau)}
\left[ a(\tau) \bar{b}(\tau) - \bar{a}(\tau) b(\tau) \right] d \tau
\end{equation}
and
\begin{equation}
\label{integral-amplitude}
|b(t)|^2 = e^{2 \gamma t} \left( |b_0|^2 - i \int_{0}^t e^{-2 \gamma \tau}
\left[ a(\tau) \bar{b}(\tau) - \bar{a}(\tau) b(\tau) \right]  d \tau \right).
\end{equation}

First, we show that the product $ab$ remains bounded as a function of $t$ for all times. Setting
$$
u := \bar{a} b + a \bar{b}, \quad v := i (a \bar{b}-\bar{a} b)
$$
and using the system of dimer equations (\ref{dimer}), we obtain
\begin{eqnarray}
\left\{ \begin{array}{l} \frac{d u}{d t} = (|b|^2-|a|^2) v, \\
 \frac{d v}{d t} = (|b|^2-|a|^2) (2 - u). \end{array} \right.
\label{dimer-phases}
\end{eqnarray}
The system (\ref{dimer-phases}) reduces to the harmonic oscillator equation
in the new temporal variable
$$
s(t) := \int_0^t (|b(t')|^2 - |a(t')|^2) dt'.
$$
Therefore, we obtain the exact solution of the system (\ref{dimer-phases}):
\begin{equation}
\label{exact-ab-variable}
\left\{ \begin{array}{l} u(t) = 2 + C_1 \cos\left[\int_0^t (|b|^2 - |a|^2) dt'\right] + C_2
\sin \left[\int_0^t (|b|^2 - |a|^2) dt'\right], \\
v(t) = -C_1 \sin\left[\int_0^t (|b|^2 - |a|^2) dt'\right] + C_2
\cos \left[\int_0^t (|b|^2 - |a|^2) dt'\right], \end{array} \right.
\end{equation}
where $C_1$ and $C_2$ are arbitrary constants, which are uniquely
defined by the initial conditions. From (\ref{exact-ab-variable}), we obtain
\begin{equation}
\label{bound-ab}
|a(t) b(t)| \leq 1 + |C_1| + |C_2|, \quad t \geq 0,
\end{equation}
hence $ab$ is a bounded function of $t$ for all times. Note that
this result also follows from the conservation law (\ref{conserved-1}) in Appendix A.

Next, we show that the component $b(t)$ remains bounded for all times if and only if
the solution of the system (\ref{dimer}) satisfies the constraint
\begin{equation}
|b_0|^2 = i \int_0^{\infty} e^{-2\gamma t} \left[ a(t) \bar{b}(t) - \bar{a}(t) b(t) \right]  dt.
\label{constraint}
\end{equation}
Indeed, since $|ab|$ is a bounded function of $t$, the integral in (\ref{integral-amplitude})
is bounded for all $t \in \mathbb{R}_+$. Therefore, if the constraint (\ref{constraint}) is not satisfied,
it follows from the integral equation (\ref{integral-amplitude}) that
the solution $|b(t)|^2$ grows like $e^{2 \gamma t}$. On the other hand, if
the constraint (\ref{constraint}) is satisfied, the integral equation (\ref{integral-amplitude})
takes the form
\begin{equation}
\label{integral-amplitude-constraint}
|b(t)|^2 = i e^{2 \gamma t} \int_{t}^{\infty} e^{-2 \gamma \tau} \left[ a(\tau) \bar{b}(\tau) - \bar{a}(\tau) b(\tau) \right]  d \tau,
\end{equation}
from which the global bound follows
\begin{equation}
\label{bounds-b}
|b(t)|^2 \leq \gamma^{-1} \sup_{t \in \mathbb{R}_+} |a(t) b(t)|, \quad t \in \mathbb{R}_+
\end{equation}
and implies that $\sup_{t \in \mathbb{R}_+} |b(t)|< \infty$. On the other hand,
it follows from the integral equation (\ref{integral-amplitude-a}) that
\begin{equation}
\label{bounds-a}
|a(t)|^2 \leq |a_0|^2 e^{-2 \gamma t} + \gamma^{-1} (1 - e^{-2 \gamma t})
\sup_{t \in \mathbb{R}_+} |a(t) b(t)|, \quad t \in \mathbb{R}_+,
\end{equation}
hence $\sup_{t \in \mathbb{R}_+} |a(t)| < \infty$. Note that bound (\ref{bounds-b}) implies
$$
\sup_{t \in \mathbb{R}_+} |b(t)| \leq \gamma^{-1} \sup_{t \in \mathbb{R}_+} |a(t)|.
$$

It remains to show that the constraint (\ref{constraint}) is satisfied for all
solutions of the system of dimer equations (\ref{dimer}) starting with small initial
data $(a_0,b_0)$. Let $\delta := \sqrt{|a_0|^2 + |b_0|^2}$ be small. We would like to show
that the amplitudes $|a|$ and $|b|$ do not grow significantly on the time scales of $t = \mathcal{O}(\delta^{-2})$.
This is achieved with the transformation of the system of dimer equations (\ref{dimer})
to the normal coordinates near the zero equilibrium point.

Using the eigenvalues and eigenvectors in (\ref{eigenvalues-dimes}) near the zero equilibrium,
we define the normal coordinates $c$ and $d$ by the transformation
\begin{equation}
\label{transform-1}
\left\{
\begin{array}{l} a = c - d (\sqrt{1 - \gamma^2} + i \gamma), \\
b = c (\sqrt{1 - \gamma^2} + i \gamma) + d. \end{array} \right.
\end{equation}
The system of dimer equations is now rewritten in coordinates $(c,d)$:
\begin{eqnarray}
\left\{ \begin{array}{l} i \dot{c} = \sqrt{1 - \gamma^2} c + (|c|^2 + 2 |d|^2) c + d^2 \bar{c} + 2i \gamma c^2 \bar{d}, \\
i \dot{d} = -\sqrt{1 - \gamma^2} c + (2 |c|^2 + |d|^2) d + c^2 \bar{d} - 2i \gamma d^2 \bar{c}. \end{array} \right.
\label{dimer-normal}
\end{eqnarray}
Removing the linear terms by the phase rotation factors,
\begin{equation}
\label{transform-2}
c(t) = C(t) e^{-i \sqrt{1 - \gamma^2} t}, \quad d(t) = d(t) e^{i \sqrt{1 - \gamma^2} t},
\end{equation}
we obtain the cubic nonlinear system
\begin{eqnarray}
\left\{ \begin{array}{l} \dot{C} = -i (|C|^2 + 2 |D|^2) C - i D^2 \bar{C} e^{4 i \sqrt{1 - \gamma^2} t}
+ 2 \gamma C^2 \bar{D} e^{- 2i \sqrt{1 - \gamma^2} t}, \\
\dot{D} = - i (2 |C|^2 + |D|^2) D - i C^2 \bar{D} e^{-4 i \sqrt{1 - \gamma^2} t}
- 2 \gamma D^2 \bar{C} e^{2i \sqrt{1 - \gamma^2} t}. \end{array} \right.
\label{dimer-cubic}
\end{eqnarray}

Since the solution exists globally, for any $T \in (0,\infty)$, let us define $\epsilon$ by
$$
\epsilon := \sup_{t \in [0,T]} \sqrt{|C(t)|^2 + |D(t)|^2}.
$$
By Gronwall's inequality, we obtain from the system (\ref{dimer-cubic}):
\begin{equation}
\label{bound-Gr}
|C(t)|^2 + |D(t)|^2 \leq (|C_0|^2 + |D_0|^2) e^{(3 + 2 \gamma) \epsilon^2 t}, \quad t \in [0,T].
\end{equation}
Since $|C_0|^2 + |D_0|^2 = \mathcal{O}(\delta^2)$, we obtain $\epsilon = \mathcal{O}(\delta)$
if $T = \mathcal{O}(\epsilon^{-2})$, that is, if $T = \mathcal{O}(\delta^{-2})$.

Finally, if the constraint (\ref{constraint}) is not satisfied, the solution $|b(t)|^2$ grows like $e^{2 \gamma t}$
and this growth on time $t = \mathcal{O}(\delta^{-2})$ contradicts the bound (\ref{bound-Gr}),
thanks to the transformations (\ref{transform-1}) and (\ref{transform-2}). Hence
the constraint (\ref{constraint}) is satisfied for all solutions of the dimer equations (\ref{dimer})
starting with small initial data $(a_0,b_0)$, consequently these solutions remain
bounded for all positive times.

\vspace{0.25cm}

{\bf Proof of R3:} Let us consider the second equation of the system (\ref{dimer-amplitudes}) rewritten as follows:
\begin{equation}
\label{second-eq-amplitude}
 \frac{d |b|^2}{d t} = 2\gamma |b|^2 - i( \bar{b} a - b \bar{a}).
\end{equation}
Let us choose the initial data $(a_0,b_0)$ to be sufficiently large so that
$$
2\gamma |b_0|^2 - i( \bar{b}_0 a_0 - b_0 \bar{a}_0) \geq 2\gamma |b_0|^2 - 2 |a_0| |b_0| \geq 2 \gamma |b_0|^2 - 2 (1 + |C_1| + |C_2|) > 0,
$$
where we have used the bound (\ref{bound-ab}) on $\sup_{t \in \mathbb{R}_+} |a(t) b(t)|$.
For instance, if $a_0 = 0$, then $C_1 = -2$, $C_2 = 0$, and $b_0$ must satisfy
the inequality $|b_0|^2 > 3 \gamma^{-1}$. Then, by the differential equation (\ref{second-eq-amplitude}), $|b(t)|^2$ will grow
and the inequality
$$
2\gamma |b(t)|^2 - i( \bar{b}(t) a(t) - b(t) \bar{a}(t)) \geq 2 \gamma |b(t)|^2 - 2 (1 + |C_1| + |C_2|) > 0,
$$
will be preserved for all positive times. By the comparison principle for differential equations,
$|b(t)|^2$ remains larger than the lower solution that grows exponentially like $e^{2 \gamma t}$.
Then, by the integral equation (\ref{integral-amplitude}), we conclude that the growth of
$|b(t)|^2$ is exactly exponential like $e^{2 \gamma t}$.

Note that even if the component $b(t)$ for the gained oscillator grows exponentially, the component
$a(t)$ for the damped oscillator remains bounded thanks to the bounds (\ref{bound-ab}) and (\ref{bounds-a}):
$$
\sup_{t \in \mathbb{R}_+} |a(t)|^2 \leq |a_0|^2 + \gamma^{-1} \sup_{t \in \mathbb{R}_+} |a(t) b(t)|
\leq |a_0|^2 + \gamma^{-1} (1 + |C_1| + |C_2|),
$$
which is only defined by the initial data $(a_0,b_0)$. Moreover, $|a(t)|^2$ must decay exponentially
as $e^{-2 \gamma t}$ to compensate the growth of $|b(t)|^2$ and to provide the uniform bound
(\ref{bound-ab}) for all positive times. Yet, while the growth of $|b(t)|^2$ is monotonic under these conditions,
according to the above argument, the decay of $|a(t)|^2$ is typically oscillatory
(see equation (\ref{xi-growth}) in  Appendix A).

It is also important to note that the above considerations
provide a simple sufficient criterion for the exponential growth, namely
\begin{equation}
\label{condition-b-0}
|b_0|^2 > \gamma^{-1} (1 + |C_1| + |C_2|).
\end{equation}
Given $a_0$ and $b_0$, coefficients $C_1$ and $C_2$ can be directly computed from
the exact solution (\ref{exact-ab-variable}). Then, if the inequality (\ref{condition-b-0}) is satisfied,
it can be immediately inferred that the amplitude $|b(t)|$ will grow indefinitely
according to $e^{2 \gamma t}$ and $|a(t)|$ will correspondingly
decay, so that their product remains bounded.

\section{Nonlinear dynamics of a finite PT-symmetric chain}

We now consider the generalization of our results to a
finite PT-symmetric chain, which is described by the PT-dNLS equation
(\ref{dnls}) for $n \in S_N := \{1,2,...,2N\}$, subject to the Dirichlet boundary conditions
$u_0 = u_{2N+1} = 0$. In the previous works \cite{Bar,Pelin3}, it was proved that
the zero equilibrium is neutrally stable for any $\gamma \in (-\gamma_N,\gamma_N)$, where
$$
\gamma_N := 2 \cos\left( \frac{\pi N}{1 + 2N} \right).
$$
When $N = 1$, this corresponds to $\gamma_1 = 1$. In what follows, we fix $\gamma \in (0,\gamma_N)$.
We are now ready to prove the three main results R1--R3 in the general case of finite $N \in \mathbb{N}$.

\vspace{0.25cm}

{\bf Proof of R1:} For any $n \in S_N$, the squared amplitude satisfies the evolution equation
\begin{eqnarray}
\frac{d |u_n|^2}{d t} = 2\gamma (-1)^n |u_n|^2 + g_n - g_{n-1}, \quad g_n := i(u_n \bar{u}_{n+1} - \bar{u}_n u_{n+1}).
\label{dimer-amplitudes-chain}
\end{eqnarray}
Adding up all equations, we obtain the balance equation for squared amplitudes
\begin{equation}
\label{balance-chain}
\frac{d}{d t} \sum_{n \in S_n} |u_n|^2 = 2\gamma \sum_{n \in S_N} (-1)^n |u_n|^2.
\end{equation}
By Gronwall' inequality, the balance equation (\ref{balance-chain}) results in
the a priori bound
\begin{equation}
\label{Gronwall-chain}
\sum_{n \in S_n} |u_n(t)|^2 \leq \left( \sum_{n \in S_n} |u_n(0)|^2 \right) e^{2 \gamma t}, \quad t \in \mathbb{R}.
\end{equation}
Bound (\ref{Gronwall-chain}) proves that the set of amplitudes $ \{ |u_n(t)| \}_{n \in S_N}$
does not blow up in a finite time.

\vspace{0.25cm}

{\bf Proof of R2:} We rewrite the differential equations
(\ref{dimer-amplitudes-chain}) in the integral form, separately for 
odd $n$
\begin{equation}
\label{integral-amplitude-chain}
|u_n(t)|^2 = |u_n(0)|^2 e^{-2 \gamma t} + \int_{0}^t e^{-2 \gamma (t-\tau)} \left[
g_n(\tau) - g_{n-1}(\tau) \right] d \tau 
\end{equation}
and even $n$
\begin{equation}
\label{integral-amplitude-chain-even}
|u_n(t)|^2 = e^{2 \gamma t} \left( |u_n(0)|^2 + \int_{0}^t e^{-2 \gamma \tau} \left[
g_n(\tau) - g_{n-1}(\tau) \right] d \tau \right).
\end{equation}
We shall now prove that the set of components $\{ g_n \}_{n \in S_N}$ remains
bounded for all times. Setting
$$
f_n := u_n \bar{u}_{n+1} + \bar{u}_n u_{n+1}, \quad g_n := i(u_n \bar{u}_{n+1} - \bar{u}_n u_{n+1})
$$
and using the PT-dNLS equation (\ref{dnls}), we obtain
\begin{eqnarray}
\left\{ \begin{array}{l} \frac{d f_n}{d t} = (|u_{n+1}|^2-|u_n|^2) g_n +
i (\bar{u}_{n-1} u_{n+1} - u_{n-1} \bar{u}_{n+1}) + i(u_n \bar{u}_{n+2} - \bar{u}_n u_{n+2}), \\
 \frac{d g_n}{d t} = (|u_{n+1}|^2-|u_n|^2) (2 - f_n) +
(\bar{u}_{n-1} u_{n+1} + u_{n-1} \bar{u}_{n+1}) - (u_n \bar{u}_{n+2} + \bar{u}_n u_{n+2}). \end{array} \right.
\label{dimer-phases-chain}
\end{eqnarray}
Using the variation of constants method, we write
\begin{equation}
\label{exact-ab-variable-chain}
\left\{ \begin{array}{l} f_n(t) = 2 + C_n(t) e^{i s_n(t)} + D_n(t) e^{-i s_n(t)}, \\
g_n(t) = i C_n(t)e^{i s_n(t)} - i D_n(t) e^{-i s_n(t)}, \end{array} \right.
\end{equation}
where
$$
s_n(t) := \int_0^t \left( |u_{n+1}(t')|^2-|u_n(t')|^2 \right) dt',
$$
and obtain the equivalent system of differential equations
\begin{eqnarray}
\label{dimer-equivalent-amplitude}
\left\{ \begin{array}{l}
\frac{d C_n}{d t} = i(u_n \bar{u}_{n+2} - u_{n-1} \bar{u}_{n+1}) e^{-i s_n}, \\
\frac{d D_n}{d t} = i (\bar{u}_{n-1} u_{n+1} - u_{n+2} \bar{u}_{n}) e^{i s_n}. \end{array} \right.
\end{eqnarray}
Since $\bar{u}_n u_{n+1} = 1 + D_n e^{-i s_n}$, we integrate the second equation
of the system (\ref{dimer-equivalent-amplitude}) and obtain
\begin{equation}
\label{u-n-u-n}
\bar{u}_n u_{n+1} = 1 + D_n(0) e^{-i s_n(t)} + i e^{-i s_n(t)} \int_0^t
(\bar{u}_{n-1} u_{n+1} - u_{n+2} \bar{u}_{n}) e^{i s_n(\tau)} d \tau.
\end{equation}
If the amplitudes $|u_n|$ or $|u_{n+1}|$ are bounded for all times,
then $\bar{u}_n u_{n+1}$ is bounded for all times from the Cauchy--Schwarz inequality 
$|g_n| \leq 2 |u_n| |u_{n+1}|$. Therefore, to conclude that $g_n$ (or equivalently, 
$\bar{u}_n u_{n+1}$) remain bounded for all other solutions, it is sufficient 
to consider the case when either $|u_n|$ or $|u_{n+1}|$ or both grow as $t \to \infty$.

Assume that either $|u_n|$ or $|u_{n+1}|$ grow as $t \to \infty$, but not both. 
Without loss of generality, we assume that $|u_n|$ grows for even $n$. 
Then, for sufficiently large $t_0 > 0$, for which $|u_{n+1}|^2 -  |u_n|^2$ is sign-definite,
we can write the integral term as follows:
\begin{equation}
\label{integration-by-parts}
i \int_{t_0}^t
(\bar{u}_{n-1} u_{n+1} - u_{n+2} \bar{u}_{n}) e^{i s_n(\tau)} d \tau =
\int_{t_0}^t
\frac{\bar{u}_{n-1} u_{n+1} - u_{n+2} \bar{u}_{n}}{|u_{n+1}|^2 -  |u_n|^2} \frac{d}{d\tau} e^{i s_n(\tau)} d \tau.
\end{equation}
Under the same assumption that $|u_n|$ grows for even $n$, 
it follows from the integral equation (\ref{integral-amplitude-chain-even}) that 
the amplitude $|u_n|$ grows as the precise exponential rate $e^{\gamma t}$. 
Similarly, $|u_{n+2}|$ may grow but is only allowed to grow at the same exponential rate. 
As a result, the integrand 
in (\ref{integration-by-parts}) before the derivative term 
converges to a constant value as $t \to \infty$ exponentially fast. Therefore, integration by parts yields a uniform 
constant bound for all times including the limit $t \to \infty$. 
In this case, we conclude that there exist positive constants $\{ G_n \}_{n \in S_N}$ such that
\begin{equation}
\label{bound-g-n}
|g_n(t)| \leq G_n, \quad t \geq 0, \quad n \in S_N,
\end{equation}
hence $g_n$ is a bounded function of $t$ for all times.

It remains to exclude the case when both $|u_n|$ or $|u_{n+1}|$ grow 
simultaneously at the same rate such that 
$|u_{n+1}|^2 - |u_n|^2 \to 0$ as $t \to \infty$. 
Adding two amplitude balance equations (\ref{dimer-amplitudes-chain}) for odd $n$, we obtain 
\begin{eqnarray}
\frac{d}{d t} \left( |u_{n+1}|^2 +  |u_n|^2 \right) = 
2\gamma (|u_{n+1}|^2 -  |u_n|^2) + g_{n+1} - g_{n-1}.
\label{technical-eq}
\end{eqnarray}
If both $|u_n|$ or $|u_{n+1}|$ grow but $|u_{n+1}|^2 - |u_n|^2 \to 0$ as $t \to \infty$, 
then (\ref{technical-eq}) implies that $|u_{n-1}|$ or $|u_{n+2}|$ grow at the same rate, 
in other words, all squared amplitudes $\{ |u_n|^2 \}_{n \in S_N}$ grow at the same rate and 
$|u_{n+1}|^2 - |u_n|^2 \to 0$ as $t \to \infty$ for all $n$. However, this clearly contradicts the balance 
equation (\ref{balance-chain}). Therefore, this case is impossible. 

By bound (\ref{bound-g-n}), functions $g_n$ are bounded for all $t$. It follows from 
the integral equation (\ref{integral-amplitude-chain}) that 
the squared amplitudes $|u_n(t)|^2$ are bounded for all odd $n$
(they correspond to the damped oscillators):
\begin{equation}
\label{bounds-damped}
|u_n(t)|^2 \leq |u_n(0)|^2 e^{-2 \gamma t} + (2\gamma)^{-1}
(1 - e^{-2 \gamma t}) \sup_{t \in \mathbb{R}_+} (|g_n(t)| + |g_{n-1}(t)|), \quad t \in \mathbb{R}_+,
\end{equation}
hence $\sup_{t \in \mathbb{R}_+} |u_n(t)| < \infty$ for odd $n$.

On the other hand, it follows from
the integral equation (\ref{integral-amplitude-chain-even}) that 
$|u_n(t)|^2$ for all even $n$,
corresponding to the gain oscillators, is bounded for all times if and only if
the solution of the PT-dNLS equation (\ref{dnls}) satisfies the constraint
\begin{equation}
|u_n(0)|^2 = \int_0^{\infty} e^{-2\gamma t} \left[ g_{n-1}(t) - g_n(t)
\right] dt, \quad \mbox{\rm for even} \;\; n \in S_N.
\label{constraint-chain}
\end{equation}
If the constraint (\ref{constraint-chain}) is satisfied,
the integral equation (\ref{integral-amplitude-chain})
takes the form
\begin{equation}
\label{integral-amplitude-constraint2}
|u_n(t)|^2 = -e^{2 \gamma t} \int_{t}^{\infty} e^{-2 \gamma \tau} \left[
g_n(\tau) - g_{n-1}(\tau) \right] d \tau,
\end{equation}
from which the global bound follows
\begin{equation}
\label{bounds-u-n}
|u_n(t)|^2 \leq (2\gamma)^{-1} \sup_{t \in \mathbb{R}_+} (|g_n(t)| + |g_{n-1}(t)|), \quad t \in \mathbb{R}_+
\end{equation}
and implies that $\sup_{t \in \mathbb{R}_+} |u_n(t)|< \infty$ for even $n$.
Using bound (\ref{bounds-damped}) and the inequality $|g_n| \leq 2 |u_n| |u_{n+1}|$, 
we find from bound (\ref{bounds-u-n}) that for all even $n$, we have
$$
\sup_{t \in \mathbb{R}_+} |u_n(t)| \leq \gamma^{-1} \sup_{t \in \mathbb{R}_+} (|u_{n-1}(t)| + |u_{n+1}(t)|).
$$

The proof that the constraints (\ref{constraint-chain}) are satisfied for all
solutions of the PT-dNLS equation (\ref{dnls}) starting with small initial
data is similar to the case of dimers. It is achieved with the transformation
of the PT-dNLS equation to normal coordinates and subsequent control
of the solution for long times that are inversely proportional to the squared
size of the small initial data. This control is contradicted to the
exponential growth of $|u_n(t)|^2$ like $e^{2 \gamma t}$ for even $n$
if the constraint (\ref{constraint-chain}) is not satisfied.

\vspace{0.25cm}

{\bf Proof of R3:} Using the balance equation (\ref{dimer-amplitudes-chain}) for even $n$
and the global bound (\ref{bound-g-n}), we choose the initial data $\{ u_n(0) \}_{n \in S_N}$
to be sufficiently large so that
$$
2\gamma |u_n(0)|^2 + g_n(0) - g_{n-1}(0) \geq 2 \gamma |u_n(0)|^2 - G_n - G_{n-1} > 0.
$$
By the differential equation (\ref{dimer-amplitudes-chain}) for even $n$,
the squared amplitude $|u_n(t)|^2$ will grow and the inequality
$$
2\gamma |u_n(t)|^2 + g_n(t) - g_{n-1}(t)  \geq 2 \gamma |u_n(t)|^2 - G_n - G_{n-1} > 0,
$$
will be preserved for all positive times. By the comparison principle for differential equations,
$|u_n(t)|^2$ for even $n$ remains larger than the lower solution that grows exponentially like $e^{2 \gamma t}$.
Then, by the integral equation (\ref{integral-amplitude-chain}), we conclude that the growth of
$|u_n(t)|^2$ for even $n$ is exactly exponential like $e^{2 \gamma t}$.
Again, $|u_n(t)|^2$ for odd $n$ must decay exponentially
as $e^{-2 \gamma t}$ to compensate the growth of $|u_n(t)|^2$ for even $n$ and to provide the uniform bound
(\ref{bound-g-n}) for all positive times. As in the case
of the dimer, the decay of the odd sites may be oscillatory,
however, if the above inequality holds, the indefinite growth
of the even sites is monotonic.

\section{Numerical illustrations for dimers and quadrimers}

\begin{figure}[tbp]
\begin{center}
\includegraphics[width=50mm,height=40mm]{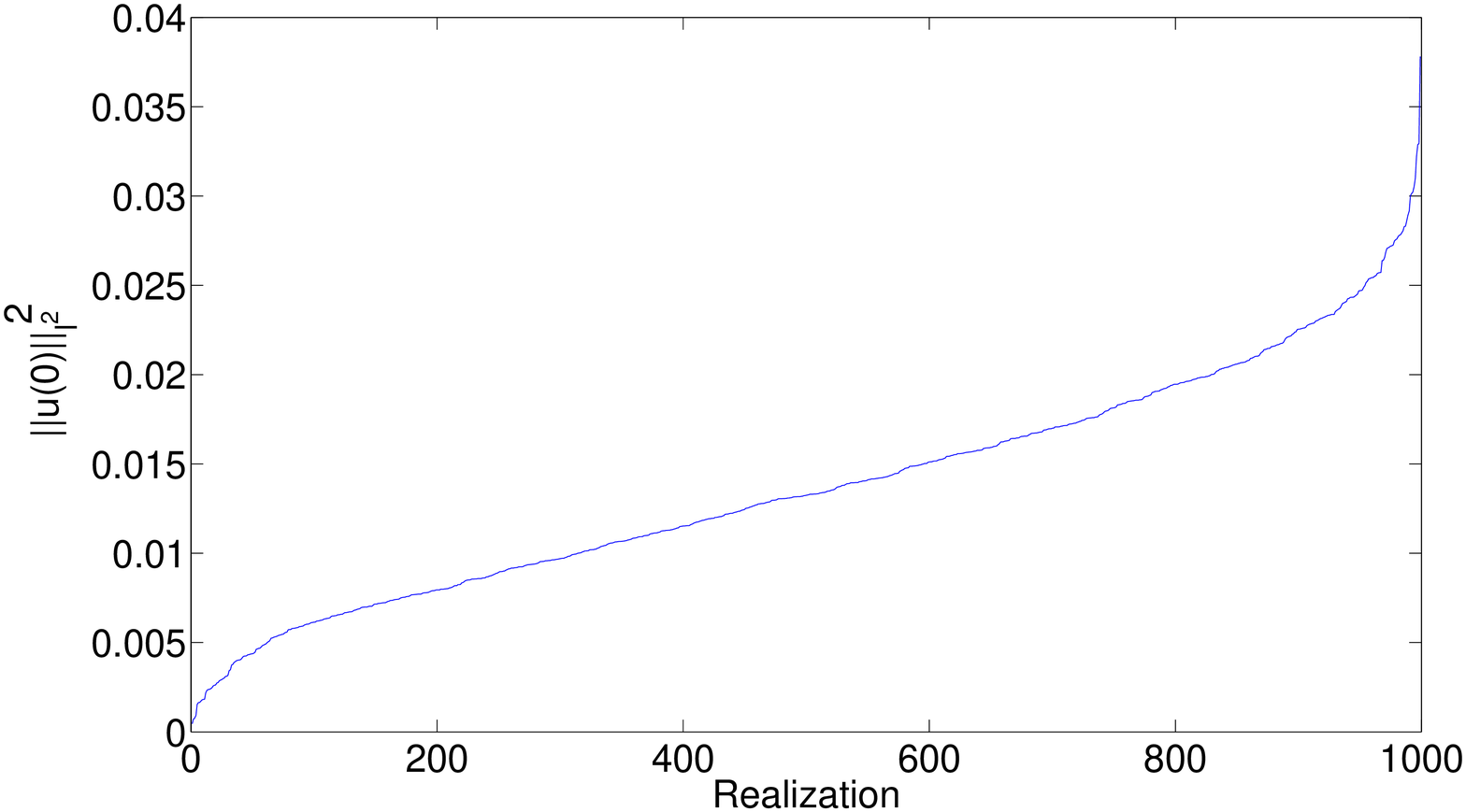}
\includegraphics[width=50mm,height=40mm]{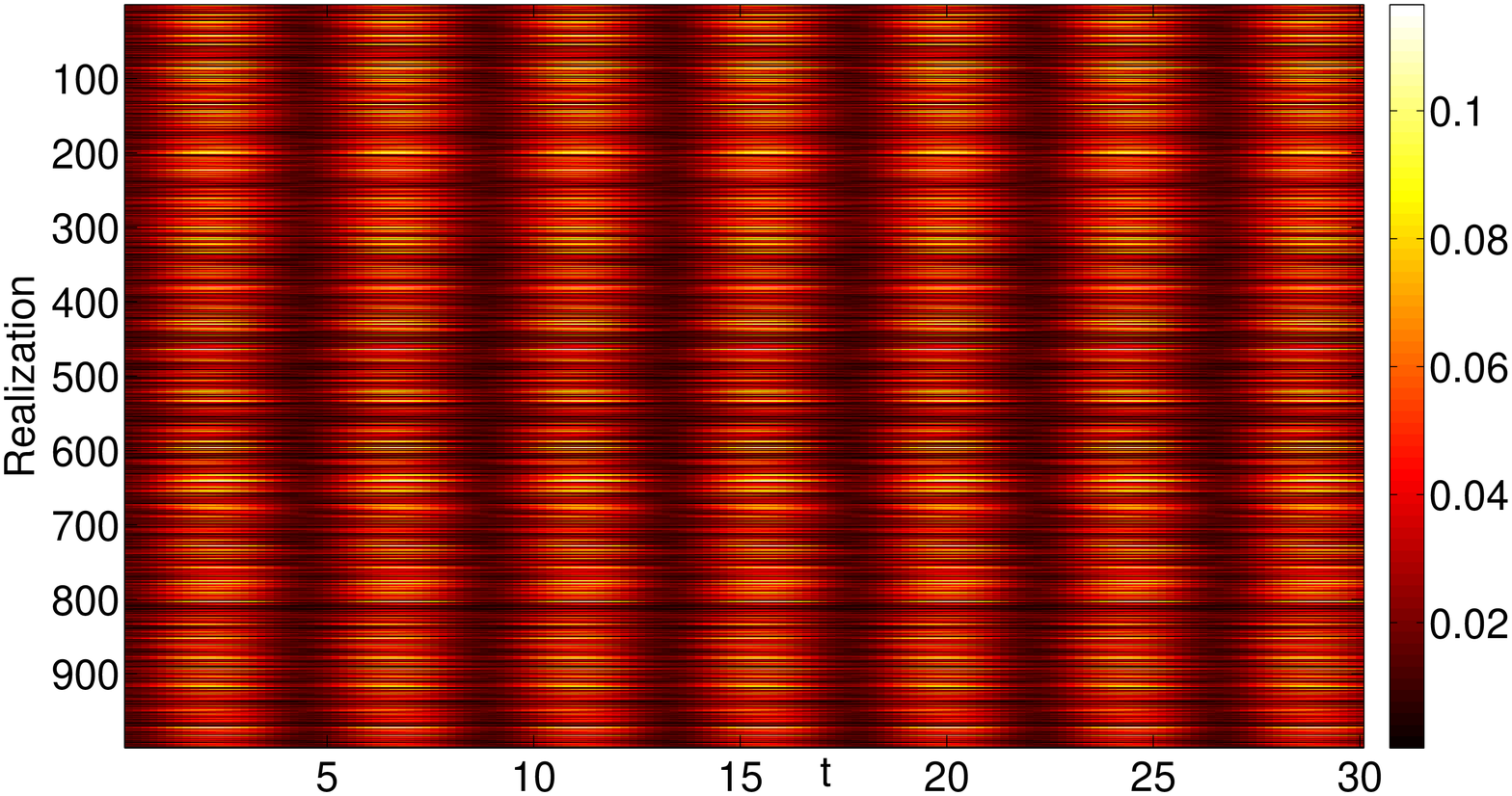}
\includegraphics[width=50mm,height=40mm]{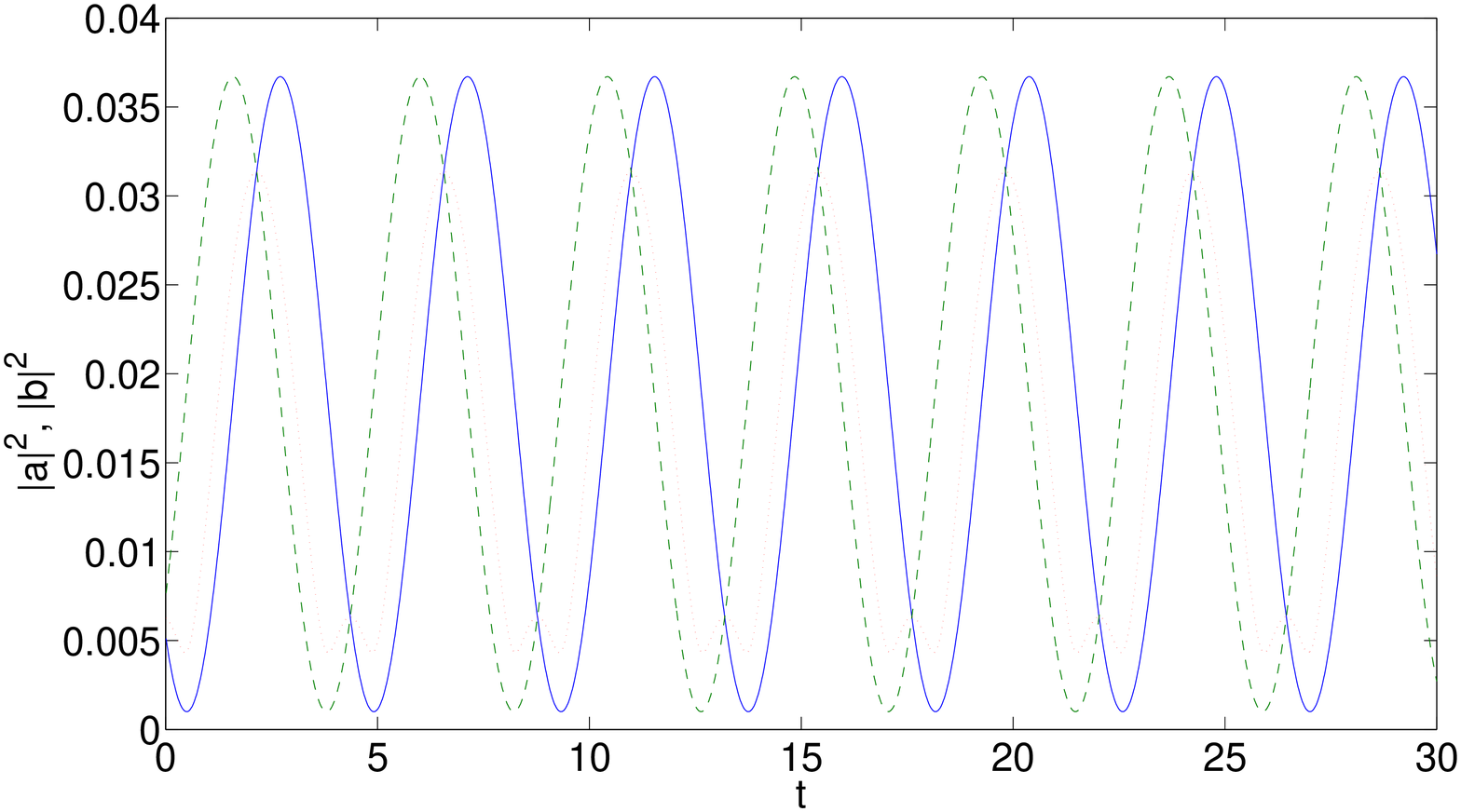}
\end{center}
\caption{The left panel shows the (sorted in increasing
value) squared numerical $l^2$
norm for our $1000$ realizations of uniformly distributed random
initial data whose real and imaginary  parts are drawn
within the interval $[0,0.1]$. The evolution of the different initial data
is shown as a colormap of the total
density $|a(t)|^2 + |b(t)|^2$ versus time (in $[0,30]$) and
the realization index in the middle panel. The bounded nature of the
results attests to the absence of indefinite growth and
the existence of oscillations. The right panel shows a typical example
of bounded oscillations for $|a(t)|^2$ (solid blue line)
and $|b(t)|^2$ (green dashed line).}
\label{tfig1}
\end{figure}

\begin{figure}[tbp]
\begin{center}
\includegraphics[width=50mm,height=40mm]{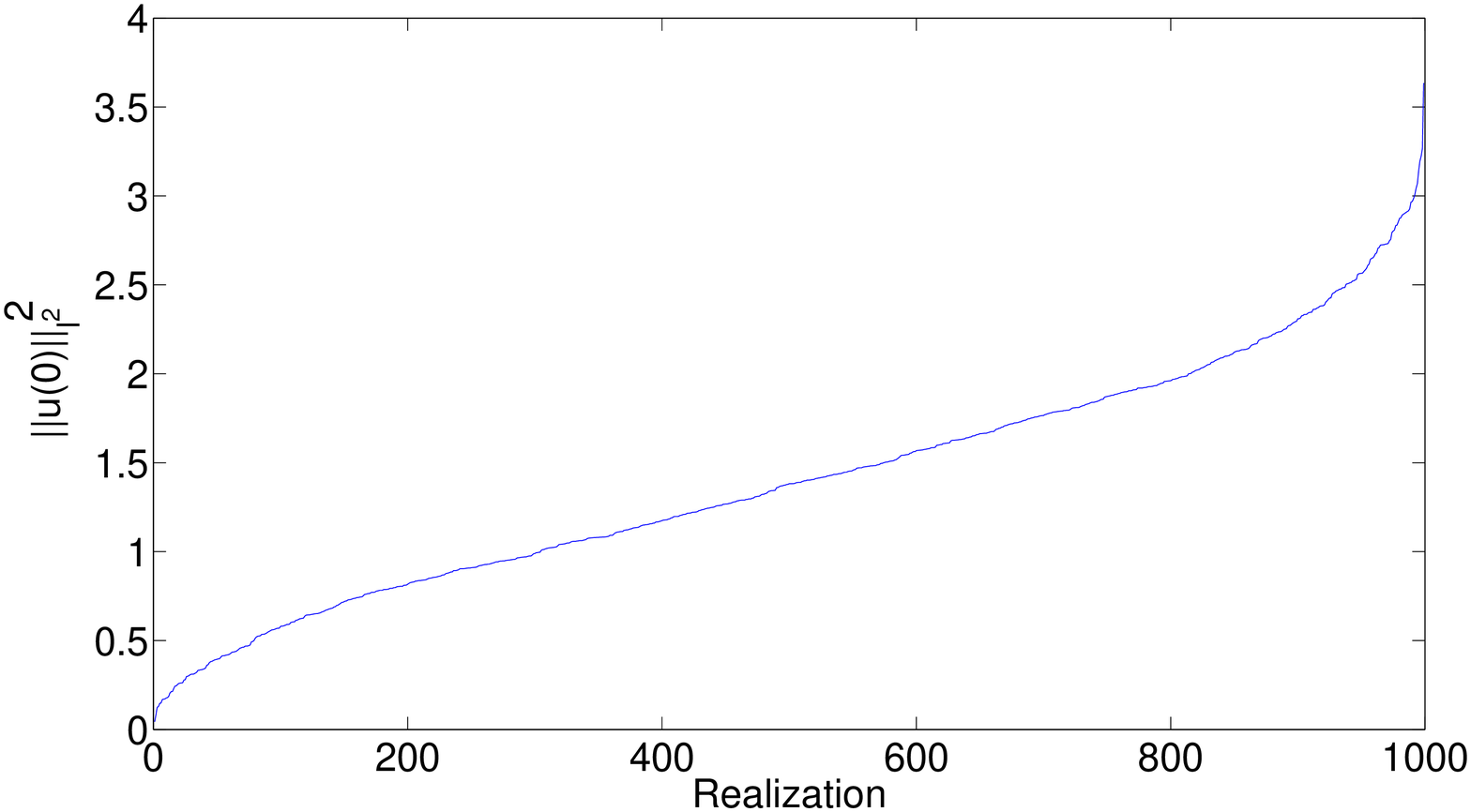}
\includegraphics[width=50mm,height=40mm]{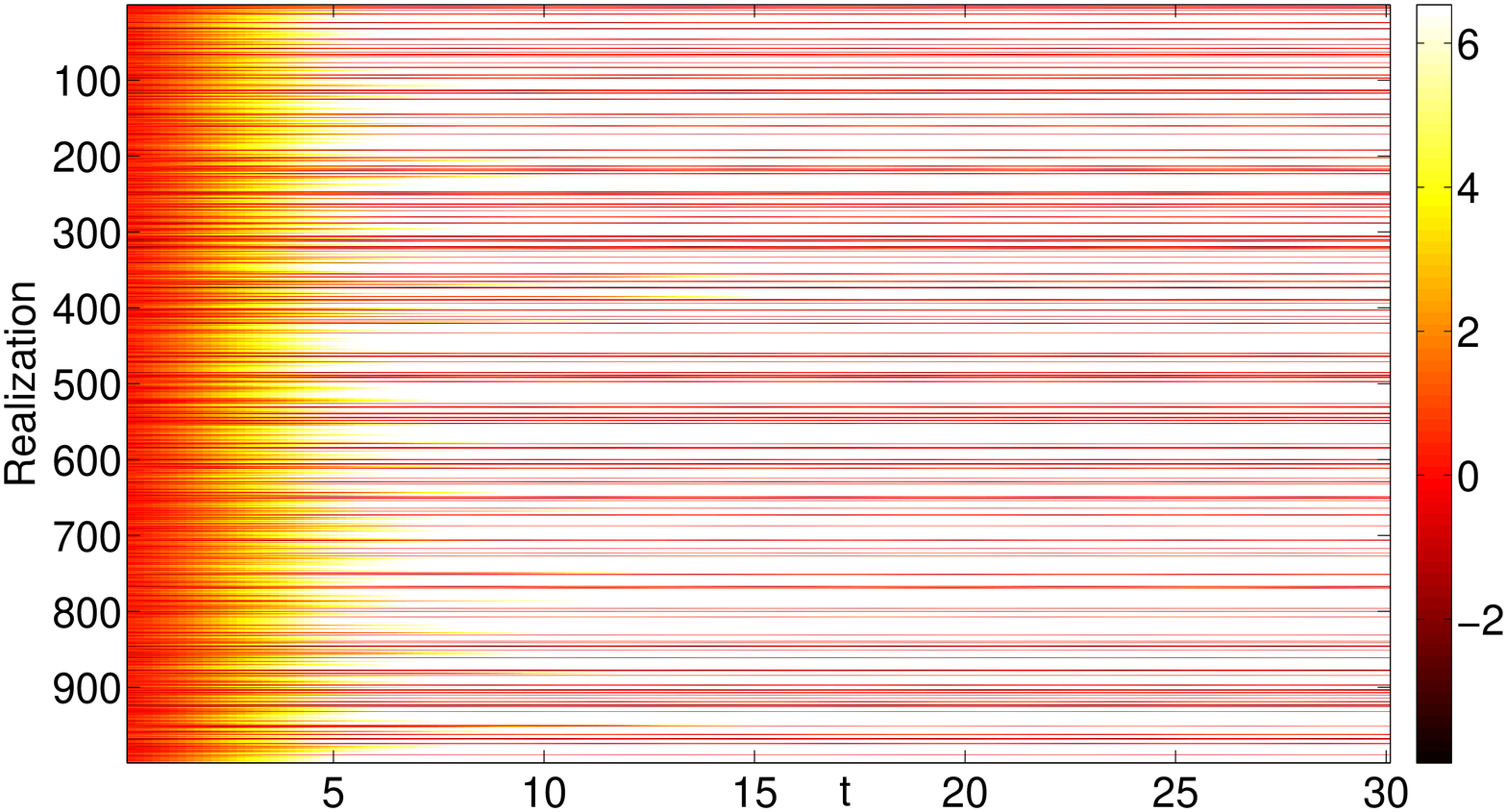} \\
\includegraphics[width=50mm,height=40mm]{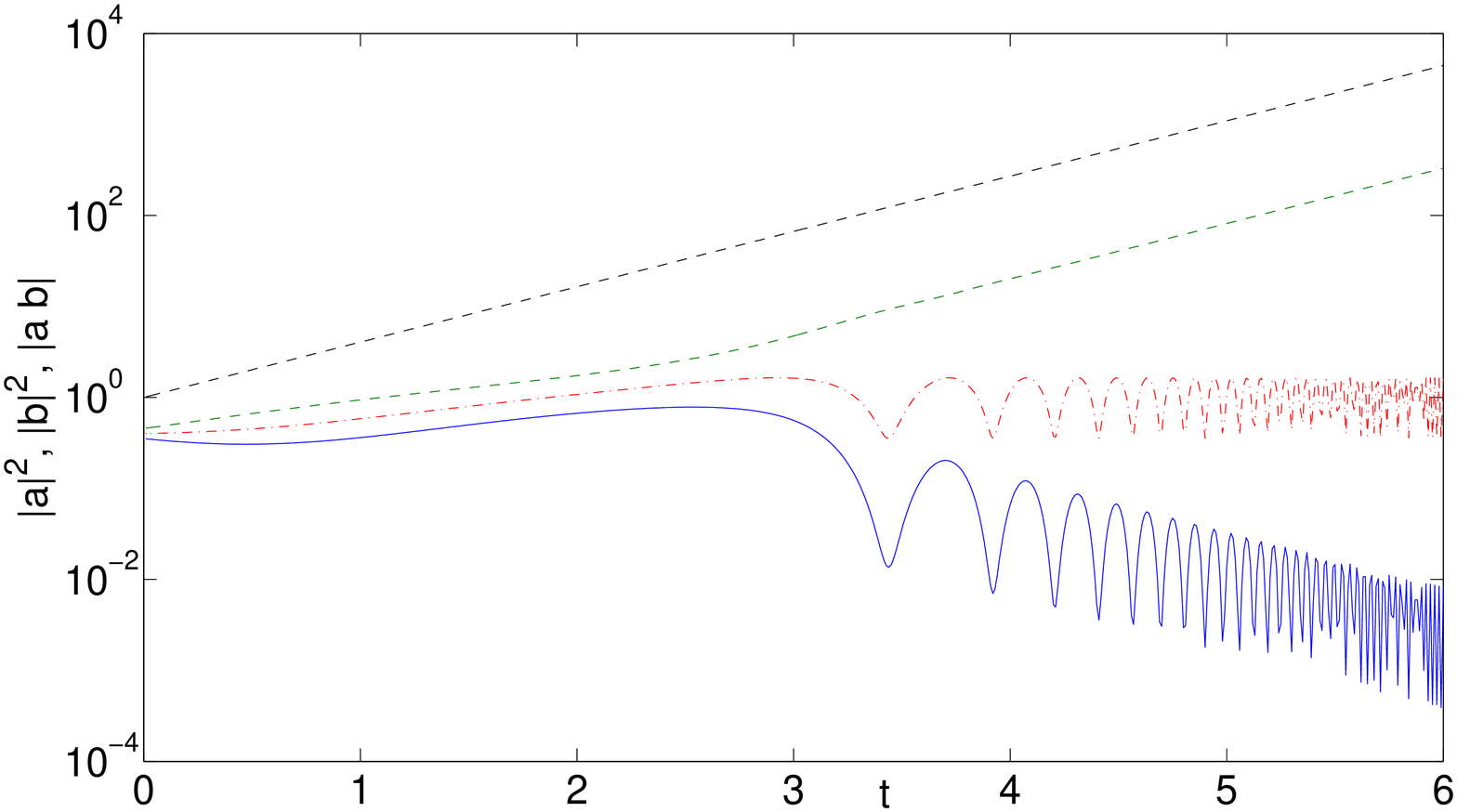}
\includegraphics[width=50mm,height=40mm]{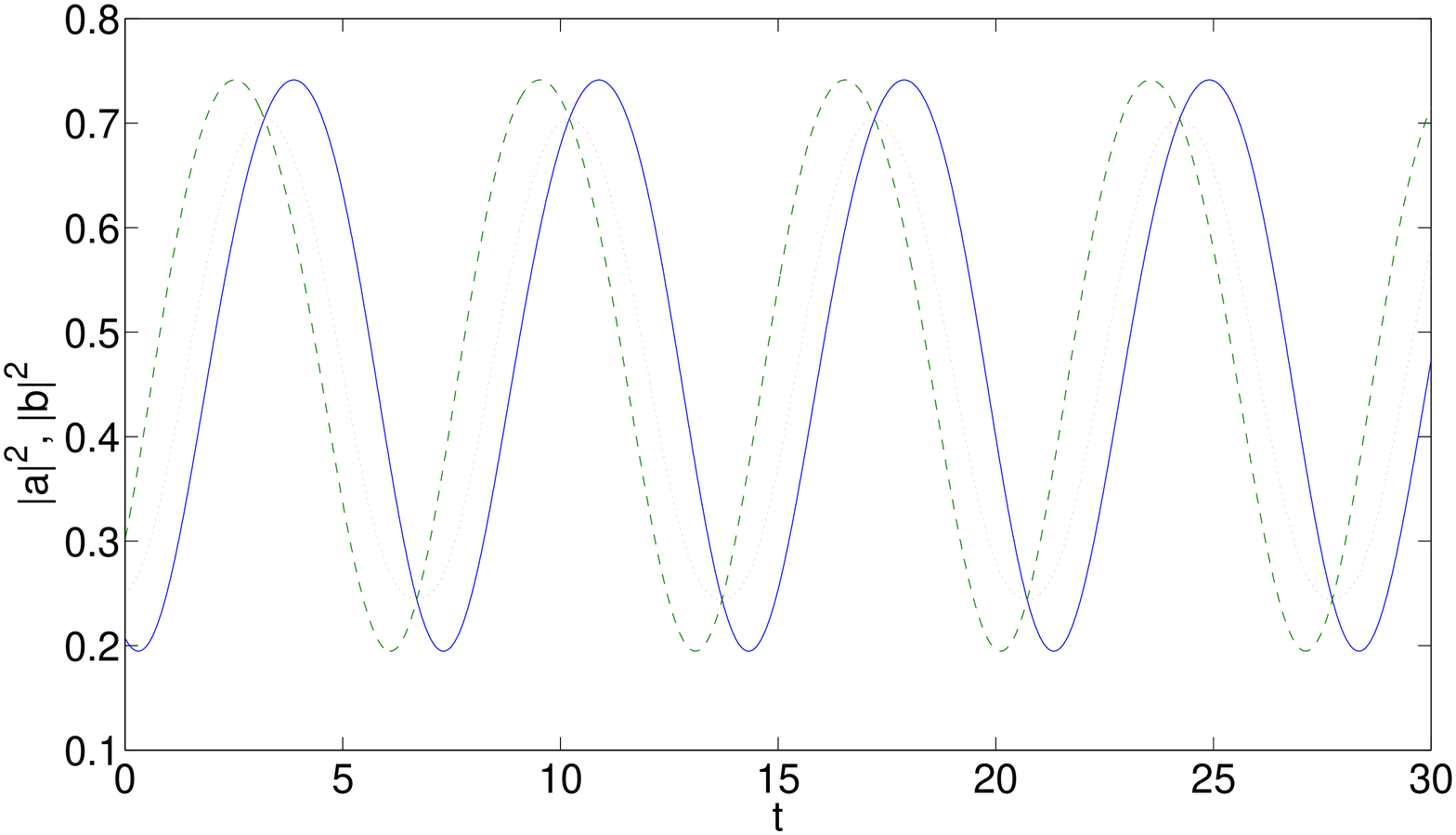}
\end{center}
\caption{The same as in Figure \ref{tfig1} but the initial amplitudes of both real
and imaginary parts are taken to be ten times larger (i.e., drawn
randomly from a uniform distribution in $[0,1]$). The top left
panel shows the (sorted) distribution of the initial square $l^2$
norms. The top right panel shows the time evolution of the different
realizations. In this case the contour plot shows 
$\log(|a(t)|^2 + |b(t)|^2)$ and the results are saturated for
large norms; i.e., for the realizations resulting in the white
regions, the evolution results in indefinite growth, while for
the ``red threads'', the evolution stays bounded for all $t$.
The bottom panels show a respective example
of the two possible scenaria (in the same form as in Fig.~\ref{tfig1}).
The only difference is that in the left panel for comparison
a dashed black line is used to depict $e^{2 \gamma t}$ in the semilog
plot, clearly indicating that the growth rate of $|b(t)|^2$
asymptotically follows the theoretical prediction. 
The bounded product $|a(t) b(t)|$ is shown also in the form of a (red) dash-dotted line.}
\label{tfig2}
\end{figure}

We now turn to a numerical illustration of the analytical results R1-R3.
The case of the dimer is considered in Figures \ref{tfig1} and~\ref{tfig2}.
For demonstration purposes, we choose $\gamma=0.7 < 1$, although
we have verified the validity of the results also for other values
of $\gamma$ in the interval $(0,1)$. 

In Fig.~\ref{tfig1}, we explore the fate
of sufficiently small initial data in connection with the statement
R2. In particular, we sample both the real and the imaginary parts
of the initial data from a uniform distribution in the interval $[0,0.1]$. 
As a result, for all $1000$ realizations considered
herein the squared initial $l^2$ norm is less than $0.04$, and hence
this case corresponds to the choice $\delta=0.2$.
We can see in the middle panel of the figure depicting the evolution
of all $1000$ realizations in time via a contour plot of
$|a(t)|^2 + |b(t)|^2$ that this quantity remains bounded (notice also
the relevant colorbar). Moreover, it typically appears to feature
oscillatory dynamics, a canonical example of which is featured
on the right panel of the figure. Hence, in accordance with the
statement R2, solutions for all sufficiently small initial
data remain bounded for the monitored times.

In Fig.~\ref{tfig2}, we explore a setting where the initial
data are sampled from an interval ten times as large, that is, 
the real and imaginary parts of $a$ and $b$ are drawn
randomly from a uniform distribution in the interval
$[0,1]$. This enables the monitoring of large initial
data settings as is clearly illustrated in the figure
(based on the corresponding initial norms). This, in turn,
leads a large fraction of the initial data to grow exponentially
over time, verifying the statement R3;
these solutions are illustrated by a saturated
white color in the contour plot of $\log(|a(t)|^2 + |b(t)|^2)$
in the top right panel of Fig.~\ref{tfig2}. On the other hand,
there are still among these $1000$ realizations ones that preserve
roughly the same (red) colormap throughout their evolution, indicating
that they correspond to bounded solutions.
An example of each of these two possibilities is illustrated
in the bottom panels of Fig.~\ref{tfig2}. On the bottom left
panel, a solution consonant with statement R3 (in that it
is associated with exponential growth) and also with statement
R1 (in that the growth happens with rate $2 \gamma$ shown for
comparison by a black dash-dotted line) is illustrated.
It is also worthwhile to note that for such solutions, 
the product $|a(t) b(t)|$ respects the bounded evolution,
whereas $|a(t)|$ decreases exponentially with some oscillations.
On the bottom right panel, an oscillatory and 
bounded solution is shown.

\begin{figure}[tbp]
\begin{center}
\includegraphics[width=50mm,height=40mm]{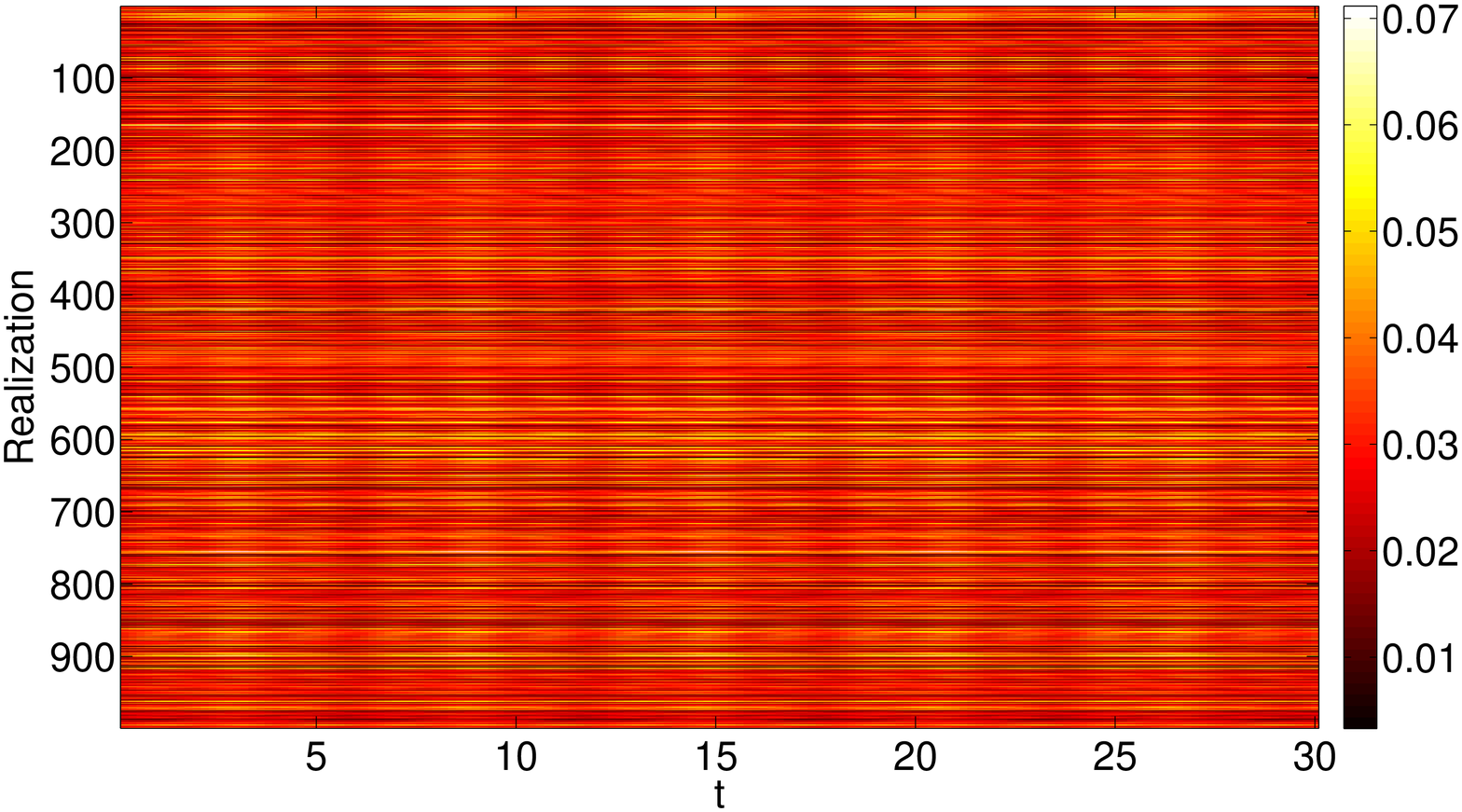}
\includegraphics[width=50mm,height=40mm]{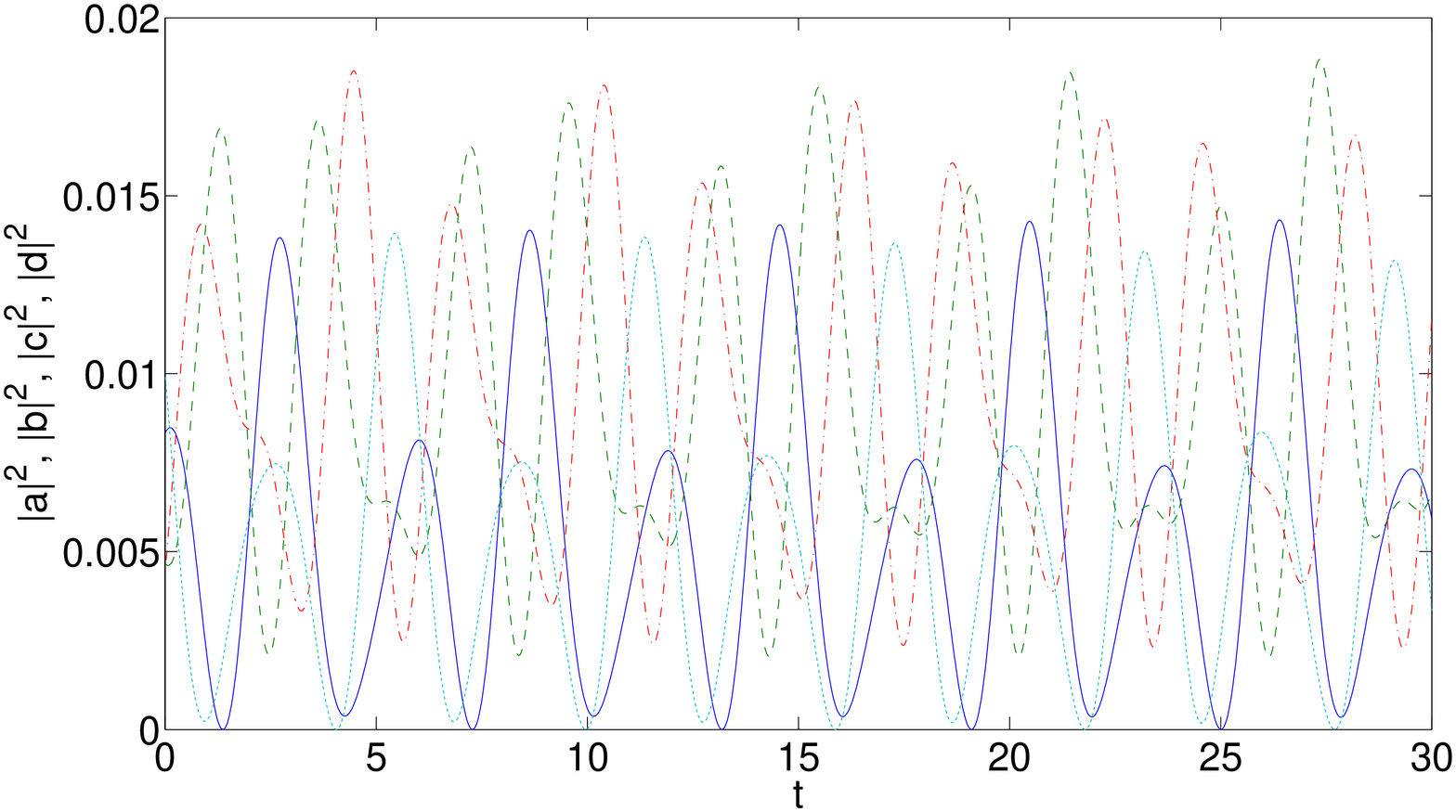}
\end{center}
\caption{Same as Figure \ref{tfig1} but for the quadrimer
with small initial data. For the contour plot evolution
of the squared $l^2$ norm, it is clear that all orbits remain
bounded (left panel). A typical example of the resulting
bounded orbit is shown in the right
panel with the blue solid and red dash-dotted lines denoting
the gain sites, while the green dashed and cyan dotted lines
correspond to the lossy ones.}
\label{tfig3}
\end{figure}

We now turn our attention to the case of quadrimers.
The case of small initial data for the quadrimer is
examined in Fig.~\ref{tfig3}. Once again, we have monitored
the evolution of $1000$ quadrimer realizations, with random
initial data chosen from a uniform distribution with both
real and imaginary parts in $[0,0.1]$. It can be clearly
seen from the colorbar that all relevant runs maintain
a norm which is bounded and follow what appears to be a nearly
periodic evolution (see, in particular, a typical case example
in the right panel of Fig.~\ref{tfig3}).

\begin{figure}[tbp]
\begin{center}
\includegraphics[width=50mm,height=40mm]{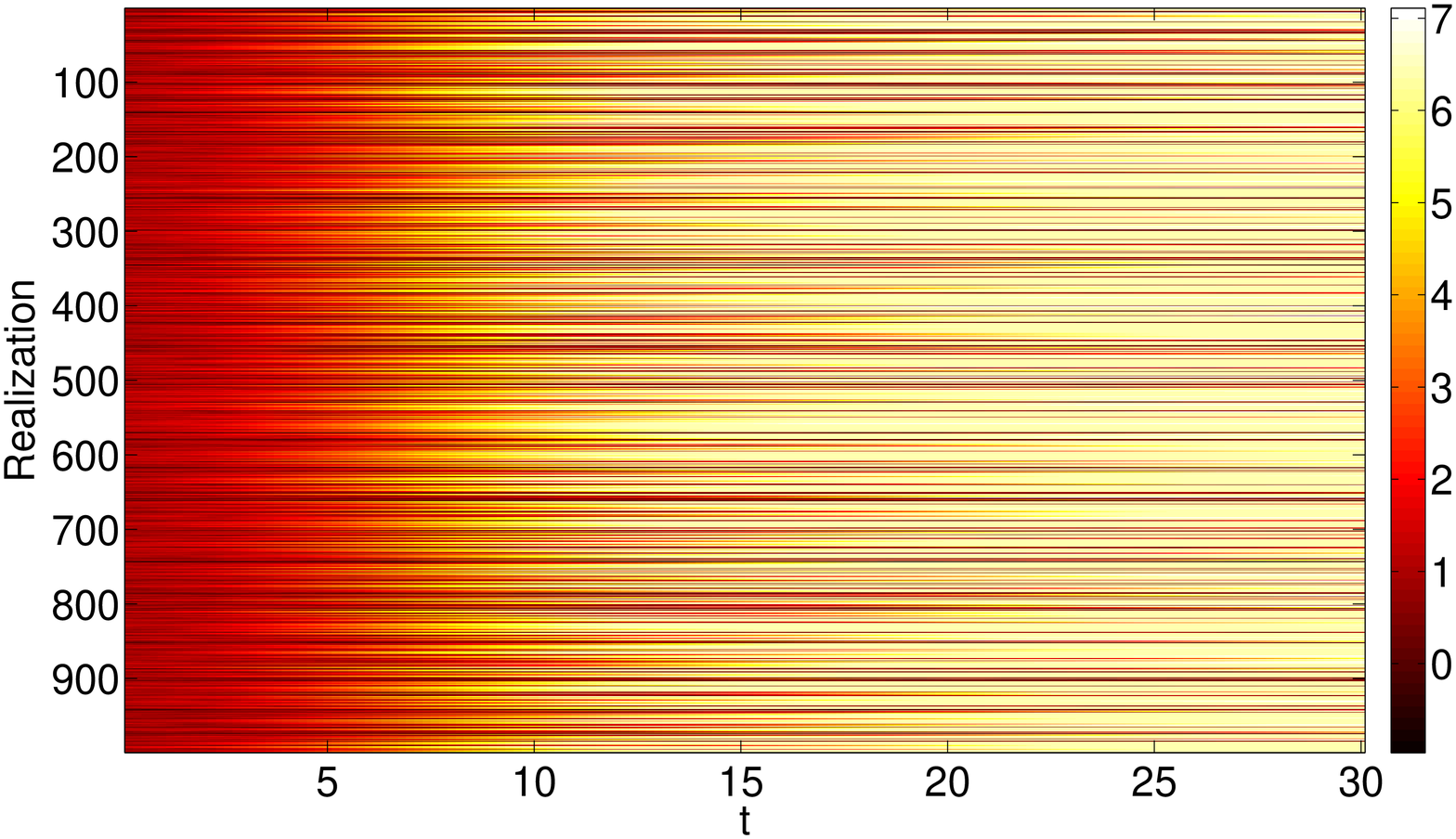}
\includegraphics[width=50mm,height=40mm]{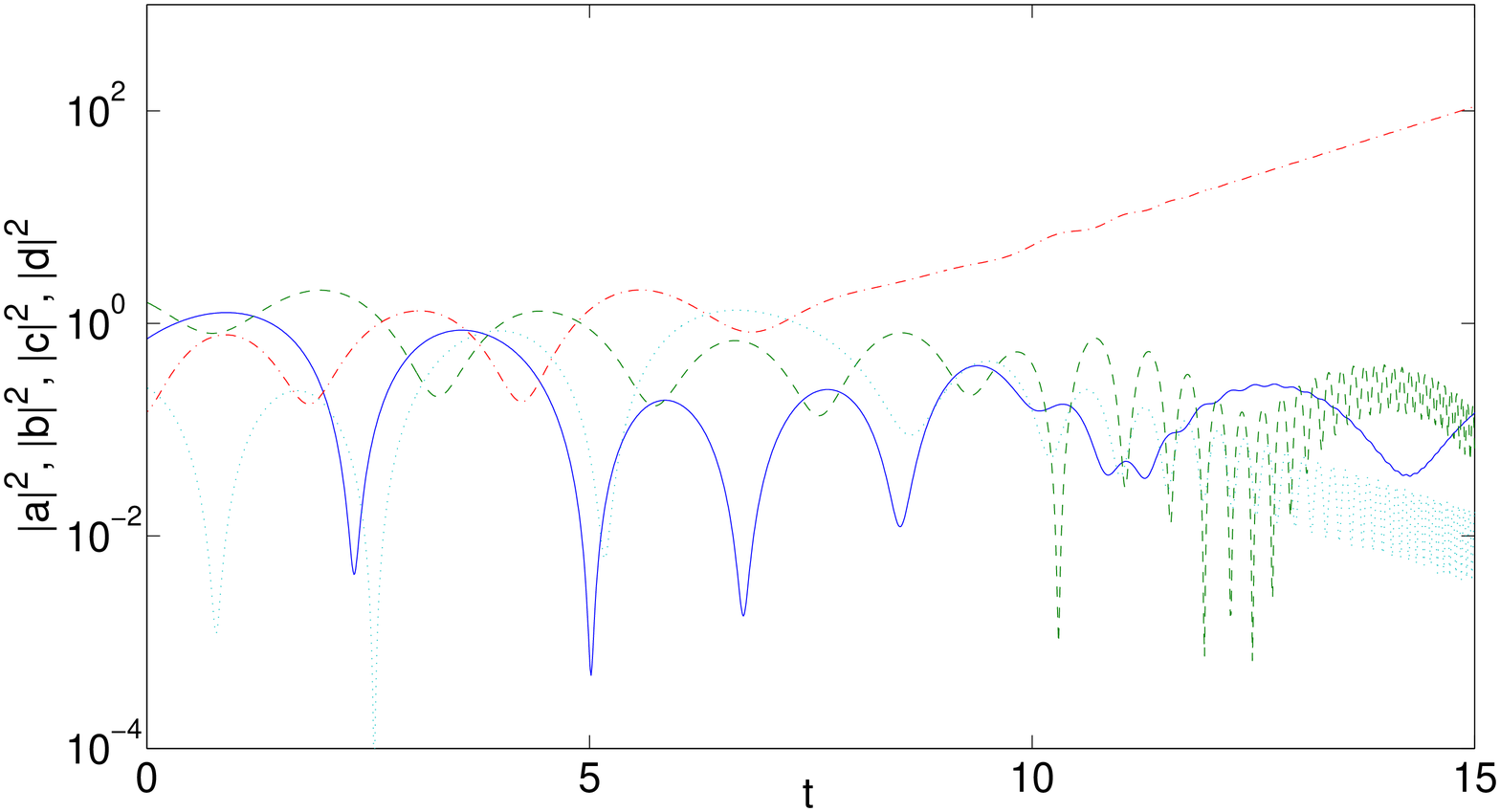}
\includegraphics[width=50mm,height=40mm]{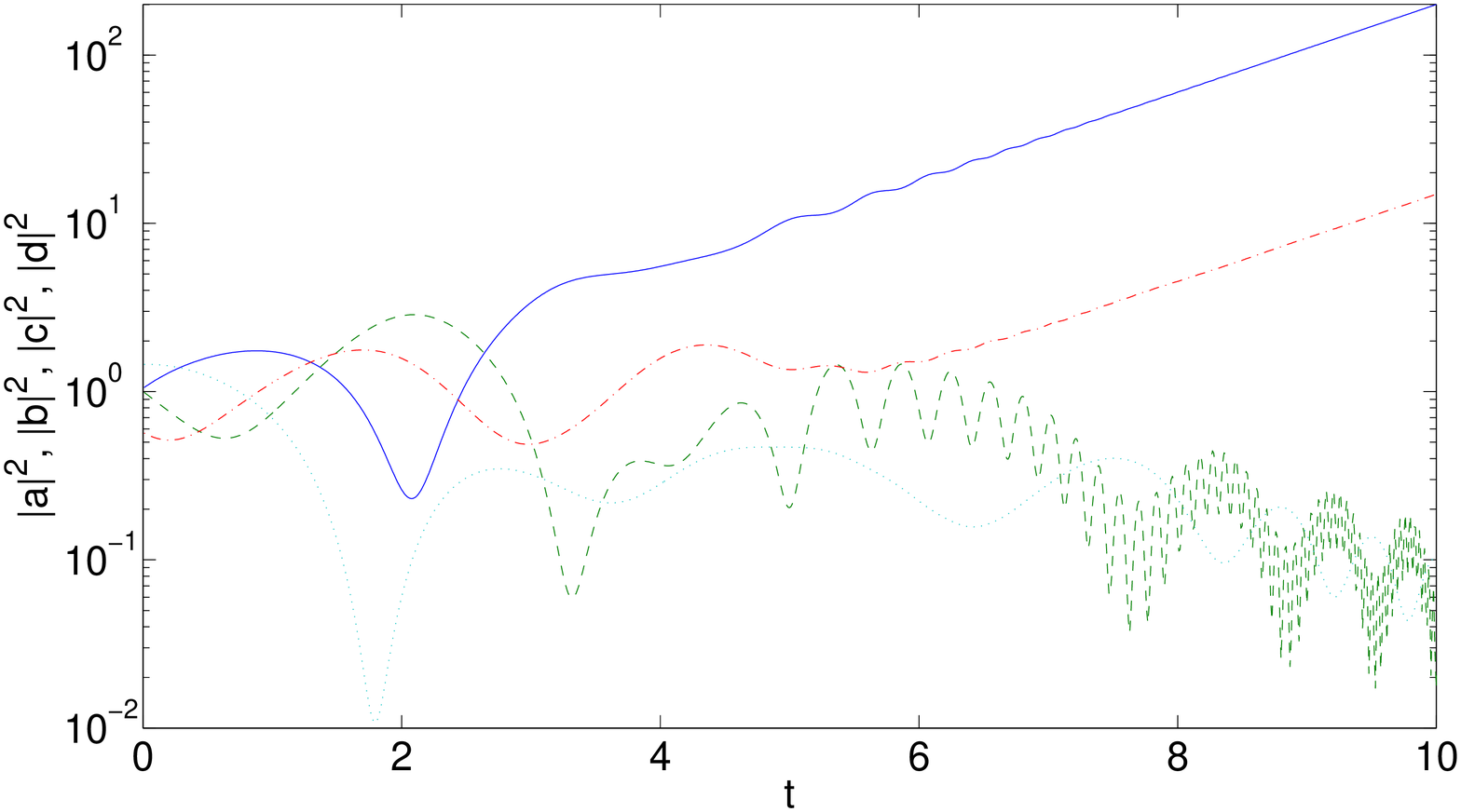}
\end{center}
\caption{Same as in Figure \ref{tfig3} for the quadrimer with large initial
data. From
the saturation of the left panel's logarithmic scale, it is clear that
most trajectories lead to indefinite growth. The middle and right panels
illustrate the two principal scenaria through which this is happening
i.e., {\it one} of the gain sites growing exponentially (middle) or {\it both}
of the gain sites growing exponentially (right).}
\label{tfig4}
\end{figure}

On the other hand, Figure \ref{tfig4} shows dynamics of
the quadrimer starting with random initial data of size ten times
larger than in Figure \ref{tfig3}. In this case, similarly to Fig.~\ref{tfig2},
we have plotted the squared $l^2$ norm of the chain in a logarithmic
contour plot, with the saturation (i.e., the yellow/faint color) indicating the
indefinite growth of most configurations. On the other hand, in this
case too, a number of solutions (the ones appearing as ``red threads''
in the left panel of Fig.~\ref{tfig4}) remain bounded.
The two additional panels of Fig.~\ref{tfig4}, middle and right,
display the two prototypical scenarios that we have observed
as being realized when indefinite growth (according to the predicted
rate of $e^{2 \gamma t}$ for the squared densities) arises for
the quadrimer configurations. In the middle panel, only one
of the two gain sites ultimately grows, while the other only
results in bounded oscillations. Nevertheless, in a number
of the relevant cases, like the one of the right panel, it is clear
that {\it both} gain sites are ultimately led towards indefinite
growth with their lossy counterparts both decaying in this case.

\section{Discussion}

In the present work, we have explored the dynamical features of
the finite PT-dNLS chains. We have centered
our exposition around three principal features of the models.

Using an a priori
estimate, based on the generalization of the $l^2$ norm (former for
$\gamma=0$) conservation law, and the Gronwall lemma, we
precluded finite time blowup for this system and offered an upper
bound on its potential growth.

Secondarily, we were able to provide
sharp bounds for the maximal possible growth of the amplitude at the sites with gain 
and the decay of the amplitudes at the sites with loss, making the interesting observation
that the product of the two stays bounded. This analysis,
in turn, permitted us to explore the fate of small data, leading
to the conclusions that solutions starting with such data always remain bounded, as may be
intuitively anticipated on the basis of the spectral stability of
the zero equilibrium (in the regime of exact PT-symmetry).

Finally, perhaps the most counter-intuitive of our findings concerned
the potential indefinite growth of solutions with large initial data {\it even in the
regime of exact PT phase}. Not only were we able to offer sufficient
conditions for such growth, but also we were able to identify its precise
rate, revealing that the growth rate indicated by the priori bound
is not merely an upper bound but a precise characterization of
the density increase. 

These results were initially presented
in systematic detail in the case of the dimer; in the latter, the complete
integrability revealed earlier in \cite{Ramezani} can be used to make
considerations far more precise, as indicated in Appendix A. However,
we intentionally chose to present proofs of our statements that
would generalize to more sites to indicate the generality of our
considerations. All of these
findings were also corroborated by systematic numerical computations,
through a large number of realizations (with random initial data).
All of these remained bounded for small initial data, as per statement
R2, but for large initial data were led to indefinite growth
both in the case of the dimer and in that of the quadrimer with
the precise rate offered by statements R1 and R3.

It will certainly be relevant to extend the present
considerations to a number of different directions.
On the one hand, it would be interesting to complement
considerations such as the ones presented herein
with more precise estimates. In particular, offering
sharper conditions for the ``separatrices'' between
indefinite growth and bounded oscillations is a particularly interesting
problem. On the other hand, the stable coherent structures of
the finite chain would naturally be anticipated to
be the centers around which the bounded motion is
organized; however, aside from the integrable case of the
dimer, this feature is not immediately transparent from
our analysis and would be quite relevant to further
explore. Finally, it is an interesting problem to consider 
higher dimensional settings, such as those in \cite{Guenter}.
The latter settings may enable the formation of more complex
phase patterns and their enhanced connectivity may modify 
the arguments devised herein.

\appendix

\section{Conservation quantities for a dimer}

We shall derive conserved quantities of the system of dimer equations (\ref{dimer}) and use them to
prove the three main results R1--R3. Note that the conserved quantities of this system were
originally reported in \cite{Ramezani} where they were introduced
with the use of the so-called Stokes variables.

Separating the amplitudes and phases of the components $(a,b)$ of the system (\ref{dimer}), we define new variables as follows:
$$
a(t) = \xi(t) e^{i \varphi(t)}, \quad b(t) = \eta(t) e^{i \varphi(t) + i \zeta(t)}.
$$
Eliminating $\varphi$, we obtain an autonomous dynamical system
for $(\xi,\eta,\zeta)$:
\begin{eqnarray}
\left\{ \begin{array}{l}
\dot{\xi} = -\gamma \xi + \eta \sin(\zeta), \\
\dot{\eta} = \gamma \eta - \xi \sin(\zeta), \\
\dot{\zeta} = (\eta^2 - \xi^2) \left[ \frac{\cos(\zeta)}{\xi \eta} - 1 \right],
\end{array} \right.
\label{3-system}
\end{eqnarray}
where the dot corresponds to the time derivative in $t$.

Equilibrium states correspond to $\xi = \eta = a$ and $\sin(\zeta) = \gamma$ with
two branches corresponding to $\cos(\zeta) = \pm \sqrt{1 - \gamma^2}$ and   an arbitrary
positive parameter $a$. Linearization at the equilibrium states results in the characteristic equation with
$$
\lambda^2 = - 4 \cos(\zeta) ( \cos(\zeta) - a^2) = \mp 4 \sqrt{1 - \gamma^2} \left( \pm \sqrt{1 - \gamma^2} - a^2\right).
$$
The upper branch is stable for $a^2 < \sqrt{1 - \gamma^2}$ and unstable for $a^2 > \sqrt{1 - \gamma^2}$.
The lower branch is stable for all $a \in \mathbb{R}_+$.

Following \cite{Ramezani}, we find two conserved quantities of the third-order system (\ref{3-system}),
which enable us to find integral curves on the phase plane $(\xi,\eta)$. The first conserved quantity is
\begin{equation}
\label{conserved-1}
E^2 := \xi^2 \eta^2 - 2 \xi \eta \cos(\zeta) + 1 = (\xi \eta - \cos(\zeta))^2 + \sin^2(\zeta) \geq 0,
\end{equation}
which is checked by direct differentiation. This conserved quantity can be used to eliminate variable $\zeta$.

The other conserved quantity follows from the balance equation
$$
\frac{d}{dt} \left[ \xi^2 \eta^2 - 2 E \sin\left( \frac{Q + \xi^2 + \eta^2}{2 \gamma} \right) \right] =
2 (\eta^2 - \xi^2) \left[ \xi \eta \sin(\zeta) - E \cos\left( \frac{Q + \xi^2 + \eta^2}{2 \gamma} \right) \right],
$$
where $Q$ is another arbitrary constant. Therefore,
\begin{equation}
\label{conserved-2}
\left\{ \begin{array}{l} E \cos\left( \frac{Q + \xi^2 + \eta^2}{2 \gamma} \right) = \xi \eta \sin(\zeta), \\
E \sin\left( \frac{Q + \xi^2 + \eta^2}{2 \gamma} \right) = \xi \eta \cos(\zeta) - 1, \end{array} \right.
\end{equation}
the compatibility condition of which is equivalent to the conserved quantity (\ref{conserved-1}).

The equilibrium states above correspond now to parametrization
\begin{equation}
\label{equilibrium}
E^2 = a^4 \mp 2 a^2 \sqrt{1 - \gamma^2} + 1, \quad
E \cos\left( \frac{Q + 2 a^2}{2 \gamma} \right) = \gamma a^2,
\end{equation}
where parameter $a \in \mathbb{R}_+$ is arbitrary and $(E,Q)$ are defined in terms of $a$.
On the other hand, we can think about parameter $Q$ as arbitrarily fixed, then
the system above fixes $a$ from the roots of a transcendental equation. This change in the point of view
is important in obtaining integral curves on the phase plane $(\xi,\eta)$.

\vspace{0.25cm}

{\bf Proof of R1:} If $(\xi,\eta)$ are initially positive and remain positive for the time span $[0,T]$,
then the third equation of system (\ref{3-system}) gives a bounded solution for $\zeta$ on $[0,T]$,
whereas the first two equations of system (\ref{3-system}) are bounded by
linear functions in $(\xi,\eta)$. This allows us to construct an upper solution
for $(\xi,\eta)$, which exists for all finite $t$ and grows
exponentially in $t$.
As a result, the time span $[0,T]$ is extended to $[0,\infty)$ provided that
the solution $(\xi,\eta)$ remain positive for all $[0,\infty)$.

To justify the positivity of $(\xi,\eta)$, we use the conserved quantity (\ref{conserved-1}).
If $E^2 \neq 1$, then $(\xi,\eta)$ cannot vanish due to the conservation of $E^2$ in (\ref{conserved-1})
and hence the local solution is extended to all $t\in [0,\infty)$. The exceptional case $E^2 = 1$
has to be treated separately.

If $E^2 = 1$ and $(\xi,\eta)$ are initially positive, then the initial data and the local solution
belong to the manifold in $\mathbb{R}^3$:
\begin{equation}
\label{manifold}
\xi \eta = 2 \cos(\zeta).
\end{equation}
Along the manifold (\ref{manifold}), system (\ref{3-system}) is rewritten in the equivalent form:
\begin{eqnarray}
\left\{ \begin{array}{l}
\dot{\xi} = -\gamma \xi + \eta \sin(\zeta), \\
\dot{\eta} = \gamma \eta - \xi \sin(\zeta), \\
\dot{\zeta} = -\frac{1}{2} (\eta^2 - \xi^2).
\end{array} \right.
\label{3-system-manifold}
\end{eqnarray}
Again, $\zeta$ is defined for all $t$, for which a solution $(\xi,\eta)$ exists, and the upper
solution for $(\xi,\eta)$ exists for all finite $t$ and grows exponentially in $t$. As a result,
the local solution is extended to $[0,\infty)$, even if $\xi$ or $\eta$ or both change sign at a finite
time instance $t_0 > 0$.

\vspace{0.25cm}

{\bf Proof of R3:} Let $x = \xi^2 + \eta^2$ and $y = \eta^2 - \xi^2$ be new dynamical variables. Excluding $\sin(\zeta)$
by using the first equation in system (\ref{conserved-2}), we transform the first two equations
of system (\ref{3-system}) to the equivalent form
\begin{eqnarray}
\left\{ \begin{array}{l}
\dot{x} = 2 \gamma y, \\
\dot{y} = 2 \gamma x - 4E \cos\left( \frac{Q + x}{2 \gamma} \right).
\end{array} \right.
\label{2-system}
\end{eqnarray}
This system, where $(Q,E)$ are given, is cast to the second-order equation
\begin{eqnarray}
\ddot{x} = 4 \gamma^2 x - 8 \gamma E \cos\left( \frac{Q + x}{2 \gamma} \right),
\label{second-order}
\end{eqnarray}
with the first integral in the form
\begin{equation}
\label{conserved-3}
I := (\dot{x})^2 - 4 \gamma^2 x^2 + 32 \gamma^2 E \sin\left( \frac{Q + x}{2 \gamma} \right).
\end{equation}
We note that $I = -16 \gamma^2 (1 + E^2)$ in connection to the conserved quantities (\ref{conserved-1}) and (\ref{conserved-2}).

Equilibrium states $(x_0,0)$ are found from the transcendental equation
\begin{equation}
\gamma x_0 = 2 E \cos\left(\frac{Q + x_0}{2 \gamma}\right),
\end{equation}
which agrees with the second equation in system (\ref{equilibrium}) by the correspondence
$x_0 = 2 a^2$. Linearization at the equilibrium states leads to the characteristic
equation
$$
\lambda^2 = 4 \gamma^2 + 4 E \sin\left(\frac{Q + x_0}{2 \gamma}\right),
$$
which is stable (unstable) if $x_0$ is a minimum (maximum) point of the effective energy
$$
V(x) := -4 \gamma^2 x^2 + 32 \gamma^2 E \sin\left( \frac{Q + x}{2 \gamma} \right).
$$
Note that $V(x) \sim -4 \gamma^2 x^2$ for large $x$, no matter what $(Q,E)$ are.
Therefore, all trajectories except for those trapped
in local minima of $V(x)$ are unbounded in the variable $x = \xi^2 + \eta^2$.

The unbounded solutions grow like $x(t) \sim e^{2 \gamma t}$
and $y(t) \sim e^{2\gamma t}$ as $t \to \infty$, which implies that $\eta(t)$ grows
exponentially in time $t$ like $\eta_{\infty} e^{\gamma t}$ as $t \to \infty$,
where $\eta_{\infty} := \lim\limits_{t \to \infty} e^{-\gamma t} \eta(t)$. From
conservation of $E^2$ in (\ref{conserved-1}), we understand that $\xi(t)$ decays exponentially in time $t$
like $e^{-\gamma t}$ as $t \to \infty$. However, the actual behavior of $\xi(t)$ is complicated as
is shown from the system (\ref{conserved-2}) after eliminating $\zeta$:
\begin{equation}
\label{xi-growth}
\xi(t) \sim \frac{1}{\eta_{\infty}}
\sqrt{1 + E^2 + 2E \sin\left( \frac{Q + \eta_{\infty}^2 e^{2\gamma t}}{2 \gamma}\right)} e^{-\gamma t}
\quad \mbox{\rm as} \quad t \to \infty.
\end{equation}
In particular, the dynamics of $\xi(t)$ features rapid oscillations and exponential decay. Notice that this is in line with the numerical observations,
as reported e.g. in the bottom left panel of Fig.~\ref{tfig2}.

\vspace{0.25cm}

{\bf Proof of R2:} Assume that the initial data satisfies
$\xi^2 + \eta^2 \leq \delta^2$ for a small parameter $\delta$. We shall first consider the dependence of conserved
quantities $(E,Q)$ as functions of $(\xi,\eta)$. From (\ref{conserved-1}), we have
\begin{equation}
\label{expansion-E}
E = \sqrt{\xi^2 \eta^2 - 2 \xi \eta \cos(\zeta) + 1} = 1 - \xi \eta \cos(\zeta) + \mathcal{O}(\xi^2 + \eta^2)^2.
\end{equation}
In what follows, we write $E = 1 + \tilde{E}$, where $\tilde{E} = \mathcal{O}(\xi^2 + \eta^2) = \mathcal{O}(\delta^2)$.
From system (\ref{conserved-2}) and expansion (\ref{expansion-E}), we have
\begin{equation}
\label{expansion-Q}
Q = 3 \pi \gamma + 2 \gamma \xi \eta \sin(\zeta) - \xi^2 - \eta^2 + \mathcal{O}(\xi^2 + \eta^2)^2.
\end{equation}
Again, we write $Q = 3 \pi \gamma + \tilde{Q}$, where $\tilde{Q} =  \mathcal{O}(\xi^2 + \eta^2) = \mathcal{O}(\delta^2)$.
We emphasize again that $(\tilde{E},\tilde{Q})$ are constants at the trajectory of the
dynamical system (\ref{3-system}) and the order of the expansion is indicated to measure its
magnitude in terms of the magnitude of the initial data.

The integral curves of the dynamical system (\ref{2-system}) are given by the first invariant (\ref{conserved-3}),
which is related to other conserved quantities by the relation $I = -16 \gamma^2 (1 + E^2)$. Substituting
our decompositions for $(E,Q)$, we obtain the integral curves in the equivalent form
\begin{equation}
\label{integral-curves}
y^2 - x^2 + 16 (1 + \tilde{E}) \sin^2\left( \frac{\tilde{Q}+x}{4 \gamma} \right) + 4 \tilde{E}^2 = 0.
\end{equation}
Expanding now the transcendental equation up to quadratic terms in the disk $0 \leq x \leq C \delta^2$
for some $C > 0$ and hiding the residual terms of the $\mathcal{O}(\delta^6)$ magnitude, we obtain the quadratic form
\begin{equation}
\label{quadratic-form}
y^2 + \frac{1}{\gamma^2} \left( \sqrt{1-\gamma^2} x + \frac{\tilde{Q}}{\sqrt{1 - \gamma^2}} \right)^2
= \frac{\tilde{Q}^2}{1 - \gamma^2} - 4 \tilde{E}^2 + \mathcal{O}(\delta^6),
\end{equation}
where $\gamma \in (0,1)$ is assumed for linear stability of the zero equilibrium. Since the quadratic form
is positive, all trajectories in the disk $0 \leq x \leq C \delta^2$ are closed curves. Therefore,
all solutions are bounded for sufficiently small initial data $(\xi,\eta)$, implying nonlinear stability
of the zero equilibrium in the dynamical system (\ref{3-system}).

\end{document}